\documentclass[pdftex,twocolumn,epjc3,natbib]{svjour3}         

\RequirePackage[T1]{fontenc}

\usepackage{xcolor}
\usepackage{cuted}
\usepackage{amsmath}
\usepackage{graphicx}
\usepackage{dcolumn}
\usepackage{multirow}
\usepackage{booktabs}
\usepackage{hyperref}
\usepackage{placeins}
\usepackage[mathscr]{euscript}
\usepackage{tabularx}
\newcolumntype{Y}{>{\centering\arraybackslash}X}

\usepackage{subcaption}

\smartqed  

\RequirePackage{graphicx}
\RequirePackage{mathptmx}      
\RequirePackage{flushend}

\journalname{
}

\newcommand{\apj}{Astrophys. J.}

\newcommand{\prd}{Phys. Rev. D}

\newcommand{\jcap}{J. Cosmol. Astropart. Phys.}

\begin{document}

\title{Precision tests of analytical tail-term approximations for radiation reaction in Schwarzschild spacetime}

\author{Bakhtinur~Juraev\thanksref{e1,addr1,addr2} \and
Arman~Tursunov\thanksref{e2,addr1} \and \\
Zden\v{e}k~Stuchlík\thanksref{addr1} \and
Martin~Kolo\v{s}\thanksref{addr1} \and
Dmitri~V.~Gal'tsov\thanksref{addr3}}
        
\thankstext{e1}{bakhtinur.juraev@gmail.com}
\thankstext{e2}{arman.tursunov@physics.slu.cz}

\institute{Research Centre for Theoretical Physics and Astrophysics, Institute of Physics, Silesian University in Opava, CZ-74601 Opava, Czech Republic\label{addr1}
          \and
National Pedagogical University of Uzbekistan, Tashkent, 100145 Uzbekistan\label{addr2}
          \and
          Faculty of Physics, Moscow State University, 119899, Moscow, Russia\label{addr3}
}

\date{Received: date / Accepted: date}

\maketitle

\begin{abstract}
We investigate the consistency and precision of approximate analytical expressions for the electromagnetic self-force acting on a charged particle in Schwarzschild spacetime endowed with weak electromagnetic fields. A fundamental requirement of relativistic particle dynamics is the preservation of the four-velocity normalization ($u^\mu u_\mu=-1$), which implies that the total self-force must remain orthogonal to the particle’s four-velocity. We introduce a covariant diagnostic based on the orthogonality condition ($u_\mu F^\mu_{\text{tail}}=0$), which provides a quantitative measure of the internal consistency of approximate tail-term models used in radiation-reaction calculations.  
We apply this diagnostic to two widely used analytical approximations for the electromagnetic tail force: the conservative component derived by Smith and Will and the dissipative component derived by Gal’tsov. The analysis is performed for several physical configurations, including pure Schwarzschild spacetime, a weakly electrically char-ged Schwarzschild black hole, and a Schwarzschild black hole immersed in a weak external magnetic field. 
We find that the conservative Smith–Will term alone leads to small but measurable deviations from the orthogonality condition, while inclusion of the dissipative Gal’tsov contribution suppresses these deviations by many orders of magnitude. For realistic radiation-reaction parameters, the violation becomes extremely small. The proposed orthogonality diagnostic offers a simple and covariant tool for validating approximate self-force models in curved spacetime and may be useful for future studies of radiation-reaction dynamics near compact objects. 
\end{abstract}

\section{Introduction} \label{sec-intro}

The motion of charged particles in curved spacetime is influenced not only by external electromagnetic and gravitational fields but also by the particle’s own radiation. The emission of electromagnetic radiation generates a self-force that acts back on the particle and modifies its trajectory. In flat spacetime this effect is described by the Lorentz–Dirac equation and its reduced \cite{Landau:1975:CTP2:} form. In curved spacetime the situation is more complex because the radiation field propagates through the curved geometry and can return to the particle due to curvature of the spacetime. As a result, the self-force acquires a nonlocal contribution, which depends on the entire past history of the particle, known as the \emph{tail term} \citep{DeW-Bre:1960:AP:,Hob:1968:AP:}. 

A general formalism describing radiation reaction in cur-ved spacetime was developed by \cite{DeW-Bre:1960:AP:} and later corrected by \cite{Hob:1968:AP:}. Later works further developed self-force theory and showed its importance for relativistic particle dynamics and gravitational-wave astrophysics \citep{Quinn-Wald:1997:PRD:,Poi-Pou-Veg:2011:LRR:}. However, the evaluation of the tail integral appearing in the self-force expression is technically challenging, since it requires knowledge of the retarded Green function along the particle worldline. For practical calculations it is therefore useful and common to employ analytical approximations for the tail term that capture the conservative and dissipative components of the self-force. 

Two classical results play an important role in this context. Smith and Will derived the conservative part of the electromagnetic self-force acting on a static charge in \linebreak Schwarzschild spacetime \citep{Smi-Wil:1980:PRD:}. Two years later, Gal'tsov obtained analytical expressions for dissipative radiation-reaction effects in the Kerr geometry \citep{Gal:1982:JPMG:}. 
These results have been widely used in studies of particle dynamics around compact objects, including analyses of radiation reaction and particle acceleration near magnetized or charged black holes \citep{Tur-Kol-Stu-Gal:2018:APJ:,San-Car-Nat:2023:PRD:,San-Car-Nat:2024:PRD:,Stuch-Kol-Tur:2024:Universe:,Jur-Stuch-Tur-Kol:2024:JCAP:,2024JHEAp..44..500S,universe7110416} with the most recent results presented by \cite{juraev2026electromagneticradiationreactionnearblack}. 
However, the extent to which these tail-term approximations remain accurate in practice has not been systematically explored, especially when they are used directly in equations of motion.  

A fundamental requirement of relativistic dynamics for a massive particle is the conservation of the normalization of the four-velocity,
\begin{equation}
u^\mu u_\mu = -1.
\end{equation}
This condition implies that the total four-force acting on the particle must remain orthogonal to its four-velocity. 
While the Lorentz force and the local radiation–reaction terms satisfy this condition exactly, approximate analytical expressions for the tail contribution may lead to small violations of the orthogonality relation. 
In particular, integrating the equations of motion with analytic tail approximations does not typically preserve the normalization relation exactly. Such deviations provide a natural diagnostic of the internal consistency of approximate self-force models.

In this work we propose a simple covariant framework for validating approximate tail-term models based on the orthogonality condition 
\begin{equation}
u_\mu F^\mu_{\text{tail}} = 0 .
\end{equation}
We interpret the magnitude of the scalar quantity $u_\mu F^\mu_{\text{tail}}$ as a measure of the precision of the adopted approximation. Using this diagnostic we investigate the accuracy of analytical tail-term approximations derived by \cite{Smi-Wil:1980:PRD:} and by \cite{Gal:1982:JPMG:}. The analysis is performed for several physically relevant configurations: a particle moving in pure Schwarzschild spacetime, a weakly electrically charged \linebreak Schwarzschild black hole, and a Schwarzschild black hole immersed in an external magnetic field. 

The study of electromagnetic radiation emitted by charged particles in curved spacetime has a long and well-established history, including analyses of synchrotron-like emission along geodesic motion~\citep{PhysRevD.8.4309}. 
The tail term can be evaluated numerically within the framework of black hole perturbation theory using the Teukolsky formalism \linebreak (see, e.g.,~\cite{2024PhRvD.109d4068G,Torres_2022}). Such computations can be carried out e.g. using the Black Hole Perturbation Toolkit~\cite{BHPToolkit}. Numerical results obtained in this way have been compared by~\cite{War-Bara:2010:PRD:} with analytical expressions derived by \linebreak ~\cite{Gal:1982:JPMG:}, and they closely match.

The structure of the paper is as follows. 
In Sec.~II we briefly review the equations of motion for a radiating charged particle and introduce the tail-term approximations used in our analysis. 
Section~III presents the numerical analysis of the orthogonality diagnostic for different dynamical regimes. 
Finally, in Section~IV we summarize and discuss the results. 


We use the spacetime signature $(-,+,+,+)$ and the $G=M = c = 4 \pi \epsilon_{0} = 1$ system of geometric units throughout the paper. In the graphics, red is used for cases with a tail term and black for without. We use the constants in the Gaussian-CGS system of units explicitly for expressions with astrophysical relevance.

\section{Orthogonality of tail-force approximations} \label{sec-dynamics}

\subsection{General equations of motion} 

The equation of motion for a radiating charged particle in curved spacetime can, in general, be written in the form
\begin{equation}
\frac{D u^{\mu}}{d\tau}
=
\mathcal{F}^{\mu}_{\rm L}
+
\mathcal{F}^{\mu}_{\rm RR}
+
\mathcal{F}^{\mu}_{\rm Hobbs}
+
\mathcal{F}^{\mu}_{\rm tail},
\label{eq_general_motion}
\end{equation}
where the left-hand side is the covariant derivative of the particle's four-velocity $u^\mu$ with respect to proper time $\tau$ along the worldline. 

The right-hand side of Eq.~(\ref{eq_general_motion}) consists of four physically distinct contributions. The term $\mathcal{F}^{\mu}_{L}$ is the Lorentz force due to an external electromagnetic field, describing the direct interaction between the particle’s charge and the ambient field. The term $\mathcal{F}^{\mu}_{\rm RR}$ is the local radiation-reaction force associated with the emission of electromagnetic radiation by the accelerated particle. The term $\mathcal{F}^{\mu}_{\rm Hobbs}$ is the local curvature-dependent part of the electromagnetic self-force, proportional to the Ricci tensor. Finally, $\mathcal{F}^{\mu}_{\rm tail}$ is the nonlocal tail term, which arises because the particle's own field propagates through curved spacetime, scatters off the geometry, and subsequently acts back on the particle. As a result, this contribution depends on the past history of the motion.

The local radiation-reaction term $\mathcal{F}^{\mu}_{\rm RR}$ contains third-order proper-time derivatives and therefore leads to the well-known unphysical pre-acceleration and runaway solutions. To avoid these pathologies, we employ the reduced-order Landau--Lifshitz (LL) equation~\citep{Landau:1975:CTP2:,Quinn-Wald:1997:PRD:,Tur-Kol-Stu-Gal:2018:APJ:}. Since the Hobbs term $\mathcal{F}^{\mu}_{\rm Hobbs}$ is proportional to the Ricci tensor, it vanishes identically in the vacuum Schwarzschild spacetime considered here. 

Thus, for a particle of mass $m$ and charge $q$ moving in an external electromagnetic field $F^{\mu\nu}$ in Schwarzschild background, the equation can be written in the following explicit form  
\begin{equation}
\begin{split}
\frac{D u^{\mu}}{d \tau}
 &= \frac{q}{m} F^{\mu}{}_{\nu} u^{\nu}
 + k \biggl(
   \frac{q}{m} F^{\mu}{}_{\nu ; \alpha} u^\nu u^\alpha
   + \frac{q^{2}}{m^{2}} F^{\mu}{}_{\nu} F^{\nu}{}_{\alpha} u^{\alpha} \\
 &\qquad\qquad
   + \frac{q^{2}}{m^{2}} F_{\alpha\beta} F^{\beta}{}_{\sigma} u^\sigma u^\mu u^{\alpha}
 \biggr)
 + \mathcal{F}^{\mu}_{\rm tail},
\label{LL}
\end{split}
\end{equation}
where
\begin{equation}
k \equiv \frac{2 q^{2}}{3 m}.
\end{equation}
The last term in Eq.~\eqref{LL} is the nonlocal tail contribution given by an integral over the particle's past worldline 
\begin{equation}
\mathcal{F}^{\mu}_{\rm tail}
=
\frac{2 q^{2}}{m} u_{\nu}
\int_{-\infty}^{\tau-0^+}
D^{[\mu} G^{\nu]}{}_{+\lambda'}
\bigl(z(\tau),z(\tau')\bigr)
u^{\lambda'}\, d\tau',
\label{tail}
\end{equation}
where $G^{\nu}{}_{+\lambda'}$ is the retarded Green function of the vector wave operator in Schwarzschild spacetime, and square brackets denote antisymmetrization. A direct evaluation of Eq.~\eqref{tail} is computationally demanding, since the Green function must be determined for all pairs of points $(z(\tau),z(\tau'))$ along the particle's past history.

\subsection{Analytic approximations to the tail term}
The exact tail force contains both conservative and dissipative contributions. In this work, we employ two analytic expressions for the tail force widely used in literature. These expressions are not exact for generic trajectories; in the parameter ranges studied below, we use them as practical estimates of the tail force. \\

\subsubsection{Smith--Will conservative radial term}
\cite{Smi-Wil:1980:PRD:} derived the analytical solution of the conservative part of the tail term in the Schwarzschild spacetime, which corresponds to the radial component of the tail term 
\begin{equation}
\mathcal{F}^{r}_{\rm tail}
=
\frac{3 k}{2 r^{3}} \sqrt{1 - \frac{2}{r}}.
\label{tail_r}
\end{equation}
The conservative part of the tail term acts as a repulsive force directed away from the black hole~\citep{Poi-Pou-Veg:2011:LRR:}. For a particle in a purely radial motion, this would be the only component of the tail term contributing to the electromagnetic self-force. 

\subsubsection{Gal’tsov dissipative terms}
\citep{Gal:1982:JPMG:} derived the dissipative part of the tail force for circular motion in Kerr spacetime. In the \linebreak Schwarzschild limit, $a=0$, the covariant components reduce to
\begin{align}
\left(\mathcal{F}_{t}\right)_{\rm tail}
   &= \frac{k \Omega_{\phi}^{2}}{r^{4}}
      \left(r^{6}\Omega_{\phi}^{2} + 4\right),
\label{tail_Gal_t} \\
\left(\mathcal{F}_{\phi}\right)_{\rm tail}
   &= - \frac{k \Omega_{\phi}}{r^{4}}
      \left(r^{6}\Omega_{\phi}^{2} + 4\right),
\label{tail_Gal_phi}
\end{align}
where $\Omega_{\phi} \equiv u^{\phi}/u^{t} = d\phi/dt$ is the azimuthal angular frequency of the particle. 
We use Eqs.~\eqref{tail_Gal_t} and~\eqref{tail_Gal_phi} as analytic estimates of the dissipative tail contribution associated with azimuthal motion. The obtained analytical solution was compared with the numerical one by \cite{War-Bara:2010:PRD:}, \linebreak and they closely match. 

\subsection{Orthogonality condition}

For any massive particle, the four-velocity satisfies the normalization condition
\begin{equation}
u_{\mu} u^{\mu} = -1.
\end{equation}
Taking the covariant derivative along the worldline yields
\begin{equation}
\frac{D}{d\tau}\left(u_{\mu} u^{\mu}\right)
=
2 u_{\mu} \frac{D u^{\mu}}{d\tau}
=
0.
\label{norm_con}
\end{equation}
Thus, the four-acceleration must remain orthogonal to the four-velocity. The explicit expression for $D u^{\mu}/d\tau$ follows from the DeWitt--Brehme--Hobbs equation. Substituting this equation into the left--hand side of Eq.~\eqref{norm_con}, we obtain
\begin{strip}
\begin{equation}
\hspace*{\fill}
    2 u_{\mu} \frac{D u^{\mu}}{d\tau}
    = \frac{2 u_{\mu}}{m} 
    \left[
        \underbrace{q F^{\mu\nu} u_{\nu}}_{\text{Lorentz force}}
        + 
        \underbrace{\frac{2 q^{2}}{3} 
        \left(
            \frac{D^{2} u^{\mu}}{d\tau^{2}} 
            + u^{\mu} u_{\nu} \frac{D^{2} u^{\nu}}{d\tau^{2}}
        \right)}_{\text{Radiation--reaction force}}
        + 
        m\,\underbrace{\mathcal{F}^{\mu}_{\text{tail}}}_{\text{Tail term}}
    \right] 
    = 2 u_{\mu} \mathcal{F}^{\mu}_{\text{tail}} .
\hspace*{\fill}
\end{equation}
\end{strip}

The first two terms are orthogonal to $u^{\mu}$,
\begin{equation}
u_{\mu} F^{\mu\nu} u_{\nu} = 0,
\qquad
u_{\mu}
\left(
\frac{D^{2} u^{\mu}}{d\tau^{2}}
+
u^{\mu} u_{\nu} \frac{D^{2} u^{\nu}}{d\tau^{2}}
\right)
= 0,
\end{equation}
and therefore Eq.~\eqref{norm_con} implies
\begin{equation}
\frac{D}{d\tau}\left(u_{\mu} u^{\mu}\right)
=
2\,u_{\mu}\mathcal{F}^{\mu}_{\rm tail}
=
0.
\end{equation}
Thus, the preservation of the normalization condition requires that the tail component of the self-force satisfy the orthogonality condition. For convenience, we introduce the following notation 
\begin{equation}
u_{\mu}\mathcal{F}^{\mu}_{\rm tail} = 0.
\label{tail_condition}
\end{equation}
We shall refer to Eq.~\eqref{tail_condition} as the \emph{orthogonality condition}. For an exact self-force, this condition is satisfied identically. For an approximate tail model, however, $u_{\mu}\mathcal{F}^{\mu}_{\rm tail}$ need not vanish exactly. Its deviation from zero therefore provides a useful diagnostic of the internal consistency of the approximation.

\begin{table*}[ht]
    \centering
    \renewcommand{\arraystretch}{1.3}
    \setlength{\tabcolsep}{10pt}
    \begin{tabular}{|cc|cc||cc|cc|}
        \hline
        \multicolumn{8}{|c|}{$\mathcal{F}^{\mu}_{\text{tail}} u_{\mu}$} \\ \cline{1-8}
        \multicolumn{4}{|c||}{Smith--Will ($\mathcal{F}^{r}_{tail}$)} 
        & \multicolumn{4}{c|}{Gal'tsov--Smith--Will ($\mathcal{F}^{r}_{tail} + \mathcal{F}^{t}_{tail} + \mathcal{F}^{\phi}_{tail}$)}  \\ \cline{1-8} 
        \multicolumn{2}{|c|}{$k=10^{-2}$} &  \multicolumn{2}{c||}{$k=10^{-19}$} &  \multicolumn{2}{c|}{$k=10^{-2}$} &  \multicolumn{2}{c|}{$k=10^{-19}$}
          \\ \cline{1-8} 
        $r_{0}=7$ & $r_{0}=70$ & $r_{0}=7$ & $r_{0}=70$ & $r_{0}=7$ & $r_{0}=70$ & $r_{0}=7$ & $r_{0}=70$ \\
        \hline \hline  
        $\sim10^{-5}$ & $\sim10^{-12}$ & $\sim10^{-22}$ & $\sim 10^{-27}$  & $\sim10^{-8}$ & $\sim10^{-14}$ & $\sim10^{-42}$ & $\sim10^{-48}$ \\
        \hline
    \end{tabular}
    \caption{Numerical values of $\mathcal{F}^{\mu}_{\text{tail}} u_{\mu}$ in the pure tail-term case without external fields for two configurations with initial positions $r_{0} = 7$ and $r_{0} = 70$. The integration proceeds until proper time $\tau_{\text{end}} = 10$ with 40 decimal digits of precision~(All numerical computations were performed in \textit{Wolfram Mathematica} using \texttt{WorkingPrecision -> 40}). 
    }
    \label{tab:tail}
\end{table*}

\section{Numerical analysis} \label{sec:numerics} 

In this section, we present numerical estimates of the precision of the adopted tail-force approximations, using the orthogonality diagnostic introduced in the previous section. We consider three representative configurations. We begin with the pure tail-force contribution in Schwarzschild spacetime, which serves as a reference case. We then extend the analysis to the cases with external Lorentz forces present, namely a weakly charged black hole and weakly magnetized black hole. In each case, we solve the equations of motion numerically and evaluate the deviations from the orthogonality condition along the trajectory.

\subsection{Tail term in pure Schwarzschild case}

In the absence of external magnetic or electric fields, the particle dynamics is governed by gravity and the electromagnetic self-force. 
Although this configuration represents the simplest physical setup, it provides a useful reference case for assessing the accuracy of the Smith--Will and Gal’tsov tail-term approximations. 
The corresponding numerical results are summarized in Table~\ref{tab:tail}. 
From these results, several conclusions can be drawn.

\begin{enumerate}
\item In the Smith–Will approximation:  
\begin{itemize}
    \item For a relatively large radiation--reaction parameter, $k = 10^{-2}$, a noticeable deviation from the orthogonality condition is observed. 
    At the initial radius $r_0 = 7$, the magnitude of the violation is $\mathcal{F}^{\mu}_{\text{tail}} u_{\mu} \sim 10^{-5}$. 
    When the particle starts motion farther from the black hole, at $r_0 = 70$, this deviation is reduced to $\sim 10^{-12}$, corresponding to an improvement of approximately seven orders of magnitude. 
    This behavior reflects the weakening of curvature effects in the asymptotically flat region, where the tail contribution becomes negligible.
    \item For a realistic radiation--reaction parameter, $k = 10^{-19}$, the violation is significantly smaller. At $r_0 = 7$, 
    we find $\mathcal{F}^{\mu}_{\text{tail}} u_{\mu} \sim 10^{-22}$, while at $r_0 = 70$ the deviation decreases further to $\sim 10^{-27}$, representing an additional suppression by about five orders of magnitude.
\end{itemize}

\item In the Gal’tsov + Smith--Will solution:
\begin{itemize}
    \item For $k = 10^{-2}$, the violation of the orthogonality condition is significantly reduced compared to the Smith--Will--only case. 
    At the initial radius $r_0 = 7$, we find $\mathcal{F}^{\mu}_{\text{tail}} u_{\mu} \sim 10^{-8}$, while at $r_0 = 70$ the deviation decreases further to $\sim 10^{-14}$.
    
    \item For the radiation--reaction parameter corresponding to an electron, $k = 10^{-19}$, the orthogonality condition is satisfied to an extremely high degree of accuracy. 
    At $r_0 = 7$, the magnitude of the deviation is $\sim 10^{-42}$, and it is further suppressed to $\sim 10^{-48}$ when the particle starts at $r_0 = 70$.
\end{itemize} 
\end{enumerate}
Overall, we conclude that in the absence of external fields the combined Smith--Will and Gal'tsov approximation provides a substantially more accurate and internally consistent description of the motion of a radiating charged particle in Schwarzschild spacetime.

\begin{table*}[p] 
    \centering
    \caption{Summary of numerical results for a weakly charged Schwarzschild black hole.}
    \label{tab:combined_all}

    \begin{subtable}[t]{\linewidth}
        \centering
        \renewcommand{\arraystretch}{1.3}
        \setlength{\tabcolsep}{10pt}
        \begin{tabular}{|c||cc|cc||cc|cc|}
            \hline
            \multirow{4}{*}{$\mathcal{J}$} & \multicolumn{8}{c|}{$\mathcal{F}^{\mu}_{\mathrm{tail}} u_{\mu}$} \\ \cline{2-9}
            & \multicolumn{4}{c||}{Repulsive case $\mathcal{J}>0$}
            & \multicolumn{4}{c|}{Attractive case $\mathcal{J}<0$} \\ \cline{2-9}
            & \multicolumn{2}{c|}{$k=10^{-2}$} & \multicolumn{2}{c||}{$k=10^{-19}$}
            & \multicolumn{2}{c|}{$k=10^{-2}$} & \multicolumn{2}{c|}{$k=10^{-19}$} \\ \cline{2-9}
            & \multicolumn{1}{c}{$r_{0}=7$}  & \multicolumn{1}{c|}{$r_{0}=70$}
            & \multicolumn{1}{c}{$r_{0}=7$}  & \multicolumn{1}{c||}{$r_{0}=70$}
            & \multicolumn{1}{c}{$r_{0}=7$}  & \multicolumn{1}{c|}{$r_{0}=70$}
            & \multicolumn{1}{c}{$r_{0}=7$}  & \multicolumn{1}{c|}{$r_{0}=70$} \\
            \hline \hline
            $10^{0}$   & $\sim10^{-6}$ & $\sim10^{-10}$ & $\sim10^{-23}$ & $\sim 10^{-29}$ & $\sim10^{-4}$ & $\sim10^{-10}$ & $\sim10^{-21}$ & $\sim10^{-27}$ \\
            $10^{-1}$  & $\sim10^{-5}$ & $\sim10^{-11}$ & $\sim10^{-22}$ & $\sim10^{-27}$  & $\sim10^{-5}$ & $\sim10^{-11}$ & $\sim10^{-22}$ & $\sim10^{-27}$ \\
            $10^{-2}$  & $\sim10^{-5}$ & $\sim10^{-12}$ & $\sim10^{-22}$ & $\sim10^{-27}$  & $\sim10^{-5}$ & $\sim10^{-12}$ & $\sim10^{-22}$ & $\sim10^{-27}$ \\
            \hline
        \end{tabular}
        \caption{
        Numerical values of $\mathcal{F}^{\mu}_{\text{tail}} u_{\mu}$ for different $\mathcal{J}$, $k$, and $r_0$. The integration proceeds until proper time $\tau_{\text{end}} = 10$ with 40 decimal digits of precision~\footnote{All numerical computations were performed in \textit{Wolfram Mathematica} using \texttt{WorkingPrecision -> 40}}.  
        }
        \label{tab:radial_electric}
    \end{subtable}

    \vspace{0.5cm} 

    \begin{subtable}[t]{\linewidth}
        \centering
        \renewcommand{\arraystretch}{1.3}
        \setlength{\tabcolsep}{12pt}
        \begin{tabular}{|c||c|c|c|c|}
            \hline
            \multirow{1}{*}{$Q\,[\mathrm{Fr}]$}
            & $F_{ C}\sim\mathcal{J}$
            & $F_{\rm tail} \sim k_{e^{-}}$
            & $F_{\rm RR1}\sim k\mathcal{J}$
            & $F_{\rm RR2}\sim k\mathcal{J}^{2}$ \\
            \hline \hline
            $10^{4}$   & $\sim10^{-6}$        & $\sim10^{-19}$ & $\sim10^{-25}$ & $\sim10^{-30}$ \\
            $10^{0}$     & $\sim10^{-10}$  & $\sim10^{-19}$ & $\sim10^{-29}$ & $\sim10^{-38}$ \\
            $10^{-2.9}$     & $\sim10^{-13}$  & $\sim10^{-19}$ & $\sim10^{-32}$ & $\sim10^{-44}$ \\
            $10^{-5}$     & $\sim10^{-15}$  & $\sim10^{-19}$ & $\sim10^{-34}$ & $\sim10^{-49}$ \\
            \hline
        \end{tabular}
        \caption{Order-of-magnitude estimates of the different forces acting on a relativistic electron near a stellar-mass black hole with mass $M = 10 M_{\odot}$. 
        All forces are normalized to the gravitational force, so that $F_{\rm G}=1$.  
        The tail force remains constant, $F_{\rm tail} \sim 10^{-19}$. } 
        \label{tab:electric_b} 
    \end{subtable}

\end{table*}

\begin{table*}[p]
    \centering
    \caption{Summary of numerical results 
    for weakly magnetized Schwarzschild black hole.}

    \begin{subtable}{1\linewidth}
        \centering
        \renewcommand{\arraystretch}{1.3}
        \setlength{\tabcolsep}{10pt}
        \begin{tabular}{|c||cc|cc||cc|cc|}
            \hline
            \multirow{4}{*}{$\mathcal{B}$} & \multicolumn{8}{c|}{$\mathcal{F}^{\mu}_{tail} u_{\mu}$ } \\ \cline{2-9}
            & \multicolumn{4}{c||}{Repulsive case $\mathcal{B}>0$} 
            & \multicolumn{4}{c|}{Attractive case $\mathcal{B}<0$}  \\ \cline{2-9} & \multicolumn{2}{c|}{$k=10^{-2}$} &  \multicolumn{2}{c||}{$k=10^{-19}$} &  \multicolumn{2}{c|}{$k=10^{-2}$} &  \multicolumn{2}{c|}{$k=10^{-19}$}
              \\ \cline{2-9} & \multicolumn{1}{c}{$r_{0}=7$} & \multicolumn{1}{c|}{$r_{0}=70$} & \multicolumn{1}{c}{$r_{0}=7$} & \multicolumn{1}{c||}{$r_{0}=70$} & \multicolumn{1}{c}{$r_{0}=7$} & \multicolumn{1}{c|}{$r_{0}=70$} & \multicolumn{1}{c}{$r_{0}=7$} & \multicolumn{1}{c|}{$r_{0}=70$}  \\
            \hline \hline
     $10^{0}$  & $\sim10^{-9}$ & $\sim10^{-15}$ & $\sim10^{-43}$ & $\sim10^{-49}$  & $\sim10^{-2}$($\tau_{end}=2$) & $\sim10^{-7}$ & $\sim10^{-36}$ & $\sim10^{-37}$ \\
     $10^{-1}$  & $\sim10^{-8}$ & $\sim10^{-14}$ & $\sim10^{-42}$ & $\sim 10^{-48}$  & $\sim10^{-6}$ & $\sim10^{-6}$ & $\sim10^{-40}$ & $\sim10^{-40}$ \\
            $10^{-2}$  & $\sim10^{-8}$ & $\sim10^{-14}$ & $\sim 10^{-42}$ & $\sim10^{-48}$  & $\sim10^{-8}$ & $\sim10^{-12}$ & $\sim10^{-42}$ & $\sim 10^{-46}$ \\
            \hline
        \end{tabular}
        \caption{
         Numerical values of $\mathcal{F}^{\mu}_{\text{tail}} u_{\mu}$ for different $\mathcal{B}$, $k$, and $r_0$. The integration proceeds until proper time $\tau_{\text{end}} = 10$ with 40 decimal digits of precision~\footnote{All numerical computations were performed in \textit{Wolfram Mathematica} using \texttt{WorkingPrecision -> 40}}. 
        }
        \label{tab:sub_a}
\end{subtable}

    \vspace{0.5cm} 

    \vspace{0.5cm} 

    \begin{subtable}{1\linewidth}
        \centering
        \renewcommand{\arraystretch}{1.3}
        \setlength{\tabcolsep}{12pt}
        \begin{tabular}{|c||c|c|c|c|}
            \hline
            \multirow{1}{*}{$B\,[\mathrm{G}]$} 
            & $F_{ L}\sim\mathcal{B}$ 
            & $F_{\rm tail} \sim k_{e^{-}}$ 
            & $F_{\rm RR1}\sim k\mathcal{B}$ 
            & $F_{\rm RR2}\sim k\mathcal{B}^{2}$ \\
            \hline \hline
            $10^{4}$      & $\sim10^{7}$   & $\sim10^{-19}$ & $\sim10^{-12}$ & $\sim10^{-5}$  \\
            $10^{0}$      & $\sim10^{3}$   & $\sim10^{-19}$ & $\sim10^{-16}$ & $\sim10^{-13}$ \\
            $10^{-2.9}$   & $\sim1$        & $\sim10^{-19}$ & $\sim10^{-19}$ & $\sim10^{-19}$ \\
            $10^{-5}$     & $\sim10^{-2}$  & $\sim10^{-19}$ & $\sim10^{-21}$ & $\sim10^{-23}$ \\
            \hline
        \end{tabular}
        \caption{Magnitudes of the various forces acting on a radiating charged particle, expressed in the dimensionless form of Eq.~(\ref{LL}), for different magnetic field strengths. The estimates correspond to a relativistic electron in the vicinity of a stellar-mass black hole with $M = 10 M_{\odot}$. The tail term is identical in all cases, $F_{\rm tail} \sim 10^{-19}$, while the gravitational force is normalized to $F_{\rm G}=1$.}
        \label{tab:magnetic_magnitude_forces}
    \end{subtable}
    
\end{table*}

\subsection{Weakly charged Schwarzschild black hole}

Next we consider a weakly electrically charged Schwarzschild black hole with charge $Q$. The weakly charged assumption implies that $Q$ is small enough that the metric remains \linebreak Schwarzschild (the $Q^2/r^2$ term in the Reissner-Nordstr\"om metric is neglected), but the term $Q/r$ in the electrostatic potential remains explicit. Such a setup is motivated by electric fields expected near black holes that can accelerate particles, as indicated by high-energy observations. The analysis follows the framework developed by \cite{Jur-Stuch-Tur-Kol:2024:JCAP:}. 

To characterize the interaction between the particle and the electric field of the black hole, we introduce the electric interaction parameter  
\begin{equation}
\mathcal{J} = \frac{qQ}{m}. 
\label{electric_int_par}
\end{equation}
Depending on the sign of $\mathcal{J}$, two regimes of motion can be distinguished. For $\mathcal{J}>0$ the Coulomb interaction is repulsive and pushes the particle away from the black hole, whereas for $\mathcal{J}<0$ it is attractive and pulls the particle inward. This setting allows us to investigate purely radial motion under the external force with the radial tail component $\mathcal{F}^{r}_{\text{tail}}$, given by the conservative Smith-Will solution \eqref{tail_r}.

We solve the equations of motion given by Eqs.~(4.6)--(4.7) from \cite{Jur-Stuch-Tur-Kol:2024:JCAP:} adding the tail-term contributions. 
The numerical results are presented in Table~\ref{tab:radial_electric}, which shows the magnitude of the orthogonality condition for different values of the electric interaction parameter $\mathcal{J}$ and the radiation--reaction parameter $k$. The following conclusions can be drawn 
\begin{enumerate}
\item Effect of the radiation-reaction parameter $k$:
\begin{itemize}

\item For $k = 10^{-2}$, the violation of the orthogonality condition ranges from $\sim 10^{-4}$ to $\sim 10^{-6}$ at $r_{0}=7$, and from $\sim 10^{-12}$ to $\sim 10^{-10}$ at $r_{0}=70$, depending on the value and sign of the electric interaction parameter.

\item For realistic $k = 10^{-19}$, the violation is significantly smaller. At $r_{0}=7$, it varies between $\sim 10^{-23}$ and $\sim 10^{-21}$, while at $r_{0}=70$ it decreases further to the range $\sim 10^{-29}$ to $\sim 10^{-27}$.

\end{itemize}

\item Effect of the electric interaction parameter $\mathcal{J}$:
\begin{itemize}

\item In the repulsive case ($\mathcal{J} > 0$), the magnitude of the orthogonality violation is generally smaller than in the attractive case ($\mathcal{J} < 0$).

\item As the magnitude of the electric interaction parameter $\mathcal{J}$ increases, the violation decreases. This behavior can be explained by the fact that the Coulomb force becomes dominant over the radiation--reaction and tail forces (see Table~\ref{tab:electric_b}), reducing the relative influence of the tail term. 

\end{itemize}

\end{enumerate}
One can conclude that, for purely radial motion where only the radial component of the tail force is included, the accuracy of the solution improves for larger values of the electric interaction parameter $\mathcal{J}$ and smaller values of the radiation--reaction parameter $k$. In addition, the solution becomes more accurate when the particle starts at larger radial distances from the black hole. This behavior reflects the rapid decrease of the tail force with distance, since its magnitude scales approximately as $\sim 1/r^{3}$.

\subsection{Weakly magnetized Schwarzschild black hole}

Next we consider a Schwarzschild black hole immersed into an external weak uniform magnetic field, 
so that the charged particle is subject to a magnetic Lorentz force in addition to gravity and radiation--reaction effects. Here we follow the setup from \cite{Tur-Kol-Stu-Gal:2018:APJ:}. To characterize the dynamics, we introduce the magnetic interaction parameter 
\begin{equation}
    \mathcal{B} = \frac{q B}{2m}, 
\end{equation}
where $B$ is the magnetic field strength. 
Analogously to the electric case, the dynamics splits into two regimes determined by the sign of 
$\mathcal{B}$. A positive value ($\mathcal{B} > 0$) corresponds to a repulsive Lorentz force that pushes the particle away from the black hole, while a negative value ($\mathcal{B} < 0$) leads to an attractive Lorentz force that pulls the particle inward.

The equations of motion are given by Eqs.~(49)--(51) from~\cite{Tur-Kol-Stu-Gal:2018:APJ:}, supplemented by the corresponding analytical expressions for the tail term. Note, that we use full set of Gal'tsov--Smith--Will components. 
Solving the equations numerically, we present the results in Table~\ref{tab:sub_a} for the various  $\mathcal{B}$, $k$, and $r_0$. The following conclusions can be drawn   
\begin{enumerate}

\item Effect of the radiation--reaction parameter $k$:
\begin{itemize}

\item For $k = 10^{-2}$, the orthogonality violation strongly depends on whether the Lorentz force is repulsive or attractive. 
In the repulsive regime ($\mathcal{B}>0$), $\mathcal{F}^{\mu}_{\text{tail}}u_{\mu}$ remains small, typically in the range $\sim 10^{-9}$ to $\sim 10^{-8}$ at $r_0=7$ and $\sim 10^{-15}$ to $\sim 10^{-14}$ at $r_0=70$. 
In the attractive regime ($\mathcal{B}<0$), the violation is generally larger, reaching $\sim 10^{-6}$ to $\sim 10^{-8}$ for moderate $\mathcal{B}$, and it can become as large as $\sim 10^{-2}$ for strong attraction (e.g., $\mathcal{B}=1$ at $r_0=7$), where the particle falls into the black hole at $\tau_{\rm end}=2$.

\item For the realistic $k = 10^{-19}$, the orthogonality condition is satisfied to extremely high accuracy in both regimes. In the repulsive case, the violation is of order $\sim 10^{-43}$--$ 10^{-42}$ at $r_0=7$ and decreases to $\sim 10^{-49}$--$10^{-48}$ at $r_0=70$. 
In the attractive case, it remains very small as well, typically between $\sim 10^{-42}$ and $\sim 10^{-36}$ at $r_0=7$ and between $\sim 10^{-46}$ and $\sim 10^{-37}$ at $r_0=70$.

\end{itemize}

\item Effect of the magnetic interaction parameter $\mathcal{B}$:
\begin{itemize}

\item Comparing the two dynamical regimes, the repulsive Lorentz force case ($\mathcal{B}>0$) generally leads to smaller values of $\mathcal{F}^{\mu}_{\text{tail}}u_{\mu}$ than the attractive case ($\mathcal{B}<0$).

\item The dependence on $\mathcal{B}$ is not strictly monotonic across both regimes. 
In the repulsive case, increasing $\mathcal{B}$ tends to keep the orthogonality violation at very small values. 

\end{itemize}

\end{enumerate}

In Table~\ref{tab:magnetic_magnitude_forces} we presents numerical results on the effect of various forces acting on the radiating particle. One can see from the table once the magnitude of the magnetic field is large, the Lorentz force becomes dominating force comparing to others. 

One can conclude that, in the magnetized case, the accuracy significantly improves compared to the charged and neutral cases. 
Unlike the purely radial-motion analysis in the previous subsection, where only the radial (conservative) tail component was included, here we account for both the conservative and dissipative parts of the tail force. 
With this more complete treatment, the orthogonality violation is reduced, by nearly twenty orders of magnitude.

\section{Conclusions}

We have investigated the accuracy of analytical approximations for the electromagnetic tail contribution to the self-force acting on a radiating charged particle in Schwarzschild spacetime endowed with weak electromagnetic fields. The central consistency requirement is the preservation of the four-velocity normalization, which implies that the total self-force must remain orthogonal to the four-velocity. In our analysis, this requirement is monitoring the value of $\mathcal{F}^{\mu}_{\text{tail}}u_{\mu}$, whose deviation from zero provides a direct quantitative measure of the precision of the adopted tail-term approximation.

In the pure-gravity case (no external electromagnetic fields), we find that the Smith--Will conservative tail term yields good accuracy in the weak-field regime, with the orthogonality violation decreasing rapidly as the initial radius increases. When the dissipative contribution derived by \linebreak  Gal'tsov is included in addition to the conservative part, the orthogonality violation is significantly suppressed. However, close to the black hole, the orthogonality violation increases becoming measurable. 

For the weakly electrically charged Schwarzschild black hole, where radial motion is considered and only the radial tail component contributes, the accuracy improves for smaller radiation--reaction parameter $k$ and for larger initial radii, reflecting the rapid decay of curvature-induced tail effects with distance. We also observe that for sufficiently large electric interaction parameter the Coulomb force dominates the dynamics, reducing the relative influence of the tail term and further decreasing the orthogonality violation.

For the weakly magnetized Schwarzschild black hole, where both conservative and dissipative tail contributions are included, the overall trends remain consistent: the approximation becomes more accurate for smaller $k$ and for larger initial radii. The sign of the magnetic interaction parameter $\mathcal{B}$ plays an important role, with attractive Lorentz force configuration exhibiting larger deviations; however, for realistic values of $k$ (e.g., the electron case) the orthogonality condition is satisfied to high precision in both for the positive and negative values of $\mathcal{B}$. 

Overall, our results show that the combined Smith–Will and Gal'tsov tail-term approximations provide a reliable description of electromagnetic self-force effects in \linebreak Schwarzschild backgrounds with and without the external Lorentz forces, yielding a more accurate representation than the Smith–Will approximation alone.

\section{Acknowledgments}
This work was supported by grants SGS/24/2024 and \linebreak IGS/27/2026 from the Silesian University in Opava. B.J. acknowledges support from the Moravian-Silesian Region Foundation under the project “Support for Talented Doctoral Students at the Silesian University in Opava 2022”. 


\end{document}